\documentclass[usegraphicx,usenatbib,useapjfonts,apj]{emulateapj}
\usepackage{numclust}
\usepackage{amssymb}

\begin{document}

\newcommand{\be}{\begin{equation}}
\newcommand{\ee}{\end{equation}}
\newcommand{\M}{\tilde{M}}
\newcommand{\C}{\mathcal{C}}
\newcommand{\rhocr}{\rho_{\mathrm{cr}}}
\newcommand{\Sigcr}{\Sigma_{\mathrm{cr}}}
\newcommand{\Mvir}{M_\mathrm{vir}}
\newcommand{\rvir}{r_\mathrm{vir}}
\newcommand{\dcr}{\delta_\mathrm{cr}}
\newcommand{\crit}{\mathrm{cr}}
\renewcommand{\sun}{\odot}
\newcommand{\Msun}{\rm M_{\odot}}
  
\title{Stacking weak lensing signals of SZ clusters to constrain cluster physics}

\author{Carolyn Sealfon,\altaffilmark{1} Licia Verde,\altaffilmark{1} Raul Jimenez\altaffilmark{1}}

\altaffiltext{1}{Department of Physics and Astronomy, University of Pennsylvania, Philadelphia, PA 19104, USA; csealfon, lverde, raulj@physics.upenn.edu}

\begin{abstract}

We show how to place constraints on cluster physics by stacking the
 weak lensing signals from multiple clusters found through the
 Sunyaev-Zeldovich (SZ) effect.  For a survey that covers about $200$
 sq. deg.  both in SZ and weak lensing observations, the slope and
 amplitude of the mass vs. SZ luminosity relation can be measured with
 few percent error for clusters at $z \sim 0.5$. This can be used to
 constrain cluster physics, such as the nature of feedback. For
 example, we can distinguish a pre-heated model from a model with a
 decreased accretion rate at more than 5$\sigma$. The power to
 discriminate among different non-gravitational processes in the ICM
 becomes even stronger if we use the central Compton parameter $y_0$,
 which could allow one to distinguish between models with pre-heating, SN
 feedback and AGN feedback, for example, at more than 5$\sigma$.
 Measurement of these scaling relations as a function of redshift
 makes it possible to directly observe e.g., the evolution of the hot gas
 in clusters.  With this approach the mass-L$_{\rm SZ}$ relation can be calibrated 
 and its uncertainties can be quantified, leading to a
 more robust determination of cosmological parameters from clusters
 surveys.  The mass-L$_{\rm SZ}$ relation calibrated in this way from a
 small area of the sky can be used to determine masses of SZ clusters
 from very large SZ-only surveys and is nicely complementary to other
 techniques proposed in the literature.

\end{abstract}


\section{Introduction}

Clusters of galaxies are powerful probes of cosmology. In particular,
the cluster number density as a function of mass and redshift can
provide unique constraints on the growth of cosmological structure and
e.g. on dark energy.

The next generation of Sunyaev-Zel'dovich (SZ) surveys, e.g., the
Atacama Cosmology
Telescope\footnote{\tt{http://www.hep.upenn.edu/act/}} (ACT) and the
South Pole Telescope\footnote{\tt{http://spt.uchicago.edu/}} (SPT),
will provide a catalog of thousands of clusters.  The flux of the SZ is proportional 
to the integral of the density
times the temperature of the hot gas in the cluster \citep{SZ80} and the flux limit of
the catalogs will approximately translate into a mass limit. However, the
actual value of the cluster's mass for a given SZ flux depends on the
details of the SZ-mass relation, which in turn is governed by cluster physics.

Recent work has shown how this relation can be obtained by using
numerical simulations (e.g. \citet{Oh03, daSilva03,Motl,Nagai}) or analytical
approximations (e.g. \citet{DosSantos01,ReidSpergel, Roy05, Ostriker05}). Numerical models find that
the SZ-mass relation is expected to be very tight, implying that
clusters masses can be directly read out from the SZ observations once
the SZ-mass relation has been calibrated. In addition, both numerical
work and analytical approximations have shown that the integrated SZ
luminosity of a cluster is relatively insensitive to input physics
used in the models, such as pre-heating and energy injection by supernovae
or Active Galactic Nuclei (e.g. \citet{Motl,Nagai,Lapi05}). The same models, however, show a larger
scatter with mass of the central SZ decrement parameter ($y_0$) and a
much stronger dependence of $y_0$ on the cluster physics.

Ideally, we would like an independent and robust measurement of the
cluster mass  from observations.
This would open the possibility to  not just  calibrate the SZ-mass
relation directly from observations, but also to test intra-cluster medium (ICM) models,
and therefore to directly probe cluster physics.
Arguably, weak gravitational lensing provides a direct way to measure
the mass of clusters. The distortions of the background galaxies due
to the cluster's  gravitational field are sensitive to the mass
along the line of sight, not just the intra-cluster gas. These distortions
can be used to reconstruct the cluster's projected mass density (e.g.,
\citet{KaiserSquires93}).

The weak lensing signal induced by galaxy clusters has been measured
for a large number of systems (e.g., \citet{Clowe05, Sheldon01} and references therein).
The amplitude of the lensing signal depends on the redshift
distribution of the faint background galaxies, which can be obtained
from photometric observations. Observations from the ground are
limited by the atmosphere: both the maximum surface density of
background galaxies and the minimum error on the shear are set by the
seeing. While observations from space could circumvent this
difficulty, there is no wide-field imager available in the near future
to do this (ACT will cover a region of a few hundred sq. deg.
and SPT 4000 sq. deg.).
  As both these factors set the
minimum cluster mass that can be directly detected from weak
lensing, a direct mass determination is possible only for
fairly massive clusters, $ \approx 10^{15} \Msun$ (e.g., \citet{Marian}).
 
Here we 
circumvent this limitation by 
taking advantage of the wide multi-wavelength coverage provided by ACT
and SPT with their planned extensive optical follow up (Southern
Cosmology Survey and Dark Energy Survey).  By combining (stacking) the weak
lensing signals from multiple clusters with roughly the
same SZ luminosity, otherwise undetectable shear signal can be
amplified, and thus an average mass determination can be achieved.

The goal of this paper is to quantify the error in the calibration of
the mass-SZ relation achievable by stacking weak lensing observations
of SZ clusters. This also will allow one to investigate cluster physics
since the scatter and amplitude in the SZ-mass relation can then be
directly compared with predictions from numerical simulations.
In addition, a well calibrated mass$-$SZ relation can lead to a more 
robust determination of cosmological parameters from cluster surveys.

Here, we will consider what can be learned by stacking
clusters from a survey with the specifications of the two-year ACT SZ
survey, assuming that the entire clean region (200 sq. deg.) is covered by optical
weak lensing observations with a seeing of $\sim 0.6$ to $0.7$ arcsec and $20$ source galaxies per square arcminute.
The outline of the paper is as follows: in section 2 we compute the
expected error in the mass estimator for the stacked clusters. In
section 3 we explore how cluster physics can be constrained with the
mass-SZ relation calibrated from the finding of section 2. We conclude
in section 4 where we also outline the possible consequences of this
approach for constraining the evolution of the hot gas in clusters and
for obtaining a more robust determination of cosmological parameters
from clusters surveys.

\section{Combining lensing signals from multiple clusters}

In ground-based experiments, the cluster weak lensing signal is
smaller than the noise for masses less than $10^{15}$ $\Msun$, due to
the low number density of background galaxies and the error with which
their sheer can be measured.  Here we illustrate how the weak lensing
signal from multiple lower--mass clusters can be combined to increase
the signal-to-noise ($S/N$) ratio. In particular, we will consider
measuring the average mass of several clusters in a given redshift bin
and mass range. We follow closely \citet{Marian} and define our
average mass estimator as the weighted sum: \be M = \sum_{d,k,s} w_{k
s, d} \gamma_T(R_k,z_s, z_d) \ee where $d$ is the index of the stacked
clusters, and $\gamma_T$ denotes the average tangential shear of
source galaxies in redshift bin $s$ in an annulus with physical radius
$R_k$ in the plane of lens $d$, and $w_{k s, d}$ are the weights,
which should be chosen e.g., to maximize $S/N$.  This is the same as
the mass estimator used by \cite{Marian}, but here it includes a
weighted sum over different clusters, and the tangential shear is in
physical rather than angular annuli.

The variance of this mass estimator is
\begin{equation}
\sigma_{M}^2 = \sum_{d,k,s} w_{ks,d}^2 \sigma^2_{\gamma_T}(R_k,z_s,z_d).
\end{equation}
The weights $w_{k s, d}$ that maximize $S/N$ are,

\be
w_{ks,d}= \left[\frac{\sum_{i,j,l} w_{ij,l} \gamma_T}{\sum_{ij,l} w_{ij,l}^2 \sigma^2_{\gamma_T}}\right] \frac{\gamma_T(R_k,z_s,z_d)}{\sigma^2_{\gamma_T}(R_k,z_s, z_d)}
\ee
The term in square brackets is constant for all $k,s$ and $d$.  Denoting this normalization constant by $\mathcal{A}$, clearly
\be
\mathcal{A}=\frac{M}{\sum_{k,s,d} \frac{\gamma_T^2}{\sigma^2_{\gamma_T}}}
\ee
which yields
\be
\label{eq:varMsum}
\sigma_{M}^2=M^2 \left(\sum_{k,s,d}\frac{\gamma_T^2(R_k,z_s,z_d)}{\sigma^2_{{\gamma_T}_{k,s,d}}}\right)^{-1}
\ee
Like \citet{Marian,BS2001}, we define
$Z(z_s,z_d) \equiv \frac{D_{ds}}{D_{s}}\Theta(z_s-z_d)$ where $\Theta$ is the
unit step function.  The shear per unit mass for a
hypothetical source at infinity is,
\be
\tilde{\gamma}_\infty(\theta_k,z_d)=\frac{\gamma_T(R_k,z_s, z_d)}{M
Z(z_s,z_d)} \ee
 The variance of $\gamma_T$ from  $N$ galaxies in annulus $k$ and redshift bin $s$ is approximated by
$\sigma_{\gamma}^2/N$ and  $\sigma_{\gamma}$ is assumed constant.
 We also  assume a constant number density $n_g$ of galaxies per
unit solid  angle on the sky, distributed in redshift  according  to
a Poisson distribution,
\be
\mathcal{P}(z_s)=\frac{1}{2 z_0^3} z_s^2 e^{\frac{-z_s}{z_0}}.
\ee
The number of galaxies in an
annulus of thickness $d\theta$ is $2 \pi n_g \sin \theta \, d\theta$, where $\theta=R/D_d$, so the number of galaxies in a physical annulus per unit redshift of a
cluster at redshift $z_d$ is $2 \pi n_g \mathcal{P}(z_s) \sin \left(\frac{R}{D_d}\right)/D_d \, d R \, dz_s$.

Taking the continuous limit of the summation over $k,s$ in Eq. \ref{eq:varMsum}, and $R \ll D_d$, we obtain
\begin{eqnarray}
\label{eq:varM}
\sigma_M^2 &=& \frac{\sigma_\gamma^2}{2 \pi n_g}   \\ \nonumber
& \times& \frac{1}{\sum_d \int_0^\infty d z_s \mathcal{P}(z_s)Z^2(z_s,z_d) \int_0^{R_\mathrm{lim}} \frac{R}{D_d^2} d R \theta \,\tilde{\gamma}_\infty^2(R,z_d)}\nonumber\\
&=& \left(\sum_d \frac{1}{\sigma^2_d}\right)^{-1},
\end{eqnarray}
where $\sigma^2_d$ is the contribution to the variance from a single lens at redshift $z_d$, as derived by \citet{Marian}. Thus, as intuitively expected, if one were to stack $N$
identical clusters (clusters with the same mass at the same redshift)
the density of background galaxies would increase by a factor $N$ and
thus the error on the measured mass would decrease by a factor
$\sqrt{N}$.

\subsection{Cluster profiles}

We assume a NFW radial cluster profile \citep{NFW},
\be
\rho(r) = \frac{\rho_s}{(r/r_s)(1+r/r_s)^2}.
\ee

We set $\rho_s$ and $r_s$ according to the clusters mass and redshift, using the  fitting formulae from \citet{Bullock01, BryanNorman}.

\be
\rho_s(z,\C)=\frac{1}{3} \delta_{\C}(\C) \Delta(z) \rhocr(z)
\ee

\be
\Delta(z)= 18 \pi^2 + 82 (\Omega(z)-1) - 39(\Omega(z)-1)^2
\ee

\be
\delta_{\mathcal{C}}=\frac{\C^3}{\ln(1+\C)-\C/(1+\C)}
\ee

\be
r_s(\Mvir,z)=\frac{\rvir(\Mvir)}{\C(\Mvir,z)}
\ee

\be
\C(\Mvir,z)=\frac{9.6} {\left(\frac{\Mvir}{M_*(z)}\right)^{0.13} (1+z)}
\ee
\be
\sigma(M_*)=\frac{1.686}{G(z)}
\ee
where $G(z)$ is the linear growth rate and $\C$ is the concentration parameter, $\Omega(z)$ is the ratio of the mean matter density to $\rhocr(z)$, $\rhocr(z)$ is the critical density of the universe at redshift $z$, and $\sigma^2(M)$ is the linear density field variance smoothed with a top-hat filter.
The mass is related to the virial radius by
\be
M \equiv \Mvir = \frac{4 \pi}{3} \Delta(z) \rhocr(z) \rvir^3.
\ee

The tangential shear is given by \citet{NFWshearpaper}: 
\be 
\gamma_{NFW}= \frac{r_s \delta_{\mathcal{C}} \rhocr}{\Sigcr} g(x) 
\ee 
\be
\Sigcr=\frac{c^2}{4 \pi G} \frac{D_s}{D_d D_{ds}}
\ee
where $x=\frac{R}{r_s}$.  Thus $\tilde{\gamma}_\infty$ becomes
\be
\tilde{\gamma}_\infty = \frac{3 G}{\Delta(D_d) c^2} \frac{D_d \delta_{\mathcal{C}}}{r_s^2
\mathcal{C}^3} g(x). 
\ee 

Following \citet{Marian}, we take the upper limit of the second integral in Eq.\ref{eq:varM}  to be $R_\mathrm{lim}=2 \rvir$, obtaining
\be
\int_0^{R_\mathrm{lim}} \frac{R}{D_d^2} d R \tilde{\gamma}_\infty^2(R) =\left(\frac{3 G}{\Delta c^2}\right)^2 \frac{\delta_{\mathcal{C}}^2}{r_s^2 \mathcal{C}^6} \int_0^{2 \mathcal{C}} x dx g^2(x)\,.
\ee

If we denote the numbers of clusters per unit redshift per unit mass per steradian as $n_c(M,z)$, and the fraction of the sky observed as $f_{\rm sky}$, then the mass estimator variance becomes,

\begin{eqnarray}
\sigma_{M}^2 = \frac{\sigma_\gamma^2}{8 \pi^2 f_{\rm sky} n_g}
\left[\int_{z_{\rm min}}^{z_{\rm max}} d z \int_{M_{\rm
min}}^{M_{\rm max}} dM n_c(M,z) \right.\nonumber\\ 
\left. \times \!\!\!\int_0^\infty \!\!d z_s
\mathcal{P}(z_s)Z^2(z_s,z) \left(\frac{3 G}{\Delta c^2}\right)^2\!\!\!\!
\frac{\delta_{\mathcal{C}}^2}{r_s^2 \mathcal{C}^6}\! \int_0^{2
\mathcal{C}} \!\!\!\!\!\! x dx g^2(x)\right]^{-1}\!\!\!\!\!.
\label{eq:varMint}
\end{eqnarray}

Lastly, we include the lognormal scatter in the concentration parameter from \citet{Bullock01}, by defining the probability distribution,

\be
P(\C', M, z)=\frac{1}{\C' \sigma_\C \sqrt{2 \pi}} e^{-\frac{(\ln \C' - \ln \C(M,z))^2}{2 \sigma_\C^2}},
\ee 

with $\sigma_\C=0.18$,
resulting in,

\begin{eqnarray}
\label{eq:varMfinal}
\sigma_{M}^2 &&= \frac{\sigma_\gamma^2}{8 \pi^2 f_{\rm sky} n_g} \left[
\int_{z_{\rm min}}^{z_{\rm max}} d z \int_{M_{\rm min}}^{M_{\rm max}} dM n_c(M,z)\right.\nonumber\\&&\times \int_{\C'}P(\C', M, z) \int_0^\infty d z_s \mathcal{P}(z_s)Z^2(z_s,z) 
\nonumber\\&&\left.\times \left(\frac{3 G}{\Delta c^2}\right)^2 \frac{\delta_{\mathcal{C}}^2}{r_s^2 \mathcal{C}'^6} \int_0^{2 \mathcal{C}'} x dx g^2(x)\right]^{-1}\;.
\end{eqnarray}

\subsubsection{Computing the available cluster number}

In order to find out the mass error achievable for a realistic survey
we need to compute how many clusters could be stacked for a given mass
bin, redshift and surveyed sky area. We use the Sheth-Tormen mass
function \citep{ShethTormen} with Eisenstein and Hu's analytical fit
\citep{EisensteinHu} for the linear mass power spectrum, and we
normalize $\sigma_8$ to get 1000 clusters above $2 \times 10^{14}$
$\Msun$ in 100 square degrees, as expected for an experiment with the
specifications of ACT.

To simplify our calculations we have found the following fitting formulae, 
\be
\sigma^2(M) \sim A_1 \left(\frac{M}{10^{14} M_\sun}\right)^{(b_1+c_1 \ln \frac{M}{10^{14} M_\sun})}
\ee
where $A_1=1.19357$, $b_1=-0.42588341$, $c_1=-0.02149738$, and
\be
\frac{d \sigma(R(M))}{d R} \sim A_2 \left(\frac{M}{10^{14} M_\sun}\right)^{(b_2+ c_2 \ln \frac{M}{10^{14} M_\sun})}
\ee
where $A_2=0.08502$, $b_2=-0.4590054$, $c_2=-0.0095691$.
The cosmological parameters we assume are $\Omega_m=0.3,
\Omega_\Lambda=0.7$ and $h=0.72$ in a $\Lambda$CDM universe
\citep{Spergel03}.  The above fitting formulae only apply for the
cosmology chosen.

\begin{table}
\begin{ruledtabular}
\begin{tabular}{cc}
\zii
\end{tabular}
\end{ruledtabular}
\caption{Mass bin sizes and number of clusters per bin for $0 < z < 0.2$ and 200 square degrees.}
\label{table:numgals1}
\end{table}

\begin{table}
\begin{ruledtabular}
\begin{tabular}{cc}
\zi
\end{tabular}
\end{ruledtabular}
\caption{Mass bin sizes and number of clusters per bin for $0.4 < z < 0.6$ and 200 square degrees.}
\label{table:numgals2}
\end{table}

\begin{table}
\begin{ruledtabular}
\begin{tabular}{cc}
\ziii
\end{tabular}
\end{ruledtabular}
\caption{Mass bin sizes and number of clusters per bin for $1.1 < z < 1.3$ and 200 square degrees.}
\label{table:numgals3}
\end{table}

\subsubsection{Accounting for the effect of the large scale structure}

The gravitational lensing signal is sensitive to all the mass along the
line of sight.  This is both a blessing and a curse: it makes lensing
the most direct technique to measure clusters masses, but it also
means that large-scale structure along the line of sight contributes
to the lensing signal and affects the mass estimates.  The effect of
local large-scale structure (filaments connecting clusters and groups)
has been studied with numerical simulations (e.g. \citet{Cen97,
Metzler99, White02}).  This effect amounts to a bias in the mass
determination and an additional scatter. While there is not yet
agreement on an estimate of the amplitude of the bias and the scatter
induced by local structures (e.g., \citet{Clowe04}), we assume here
that the bias can be accurately calibrated with numerical simulations
and that the residual scatter is negligible compared to the weak
lensing mass error. While  this assumption is expected to hold in most of the regime considered, it may break down  at the high end of
our mass range, thus more investigation is needed. This does not affect the results presented here significantly.
In addition, the SZ-mass relation is expected to be
affected less by these structures than the mass determination alone. In fact 
both weak-lensing mass and SZ signal are projections along the line of
sight, the former of mass and the latter of temperature-weighted
gas mass, and thus they are affected by these structures in approximately
the same way.

The effect of distant (not correlated with the cluster) large-scale
structure is an important source of error in the mass determination of
a single object, and it cannot be neglected.  \citet{Hoekstra2001}
showed that these large-scale structures introduce a noise in the
cluster mass but not a bias.  \citet{Hoekstra2001, White02,Hoekstra02}
quantify the effect and showed that it is roughly redshift independent
for $z \gtrsim 0.1$ and that it amounts to roughly doubling the error
estimate obtained without accounting for the large-scale structure
effect. \citet{Dodelson03} showed how to somewhat correct for this
effect, using the fact that the large-scale structure noise produces a
signal which is correlated over many pixels.  This can significantly
reduce the resulting uncertainties.  Here we consider two cases: in
the pessimistic case following \citet{Hoekstra02} we double the errors
estimated as in \S 2, in the optimistic case following
\citet{Dodelson03} we increase the errors of \S 2 by 50\%. In the text
and figures we report the pessimistic case (in the optimistic case the
errors will be reduced by a factor $1.15$), but in the tables we
report both.

\section{Constraining cluster physics with mass-SZ scaling relations}

Here we will consider two quantities as a function of cluster mass and
redshift: the total (integrated over the projected cluster area of a
cluster) SZ luminosity, $L_{sz}$, and the central Compton parameter,
$y_0$.  Both these quantities are expected to depend on clusters mass
approximately as power laws, with some intrinsic scatter. The details
of the relation between mass and either $L_{sz}$ or $y_0$ depend on
the physics of the Intra Cluster Medium (ICM). It is customary to
parameterize these relation over a range of masses as power laws with
two free parameters (an amplitude and a slope of a linear fit in
log-log space) and study how these parameters are expected to change
for different cluster physics. Turning the argument around by studying
the effect of cluster physics on two parameters one is basically
describing cluster physics with a two parameters model.  Here we
estimate how well these two parameters could be measured: using the
errors calculated for the mass bins in the previous section, we
estimate the error on the slope and amplitude of the mass-$L_{sz}$
scaling and the mass-$y_0$ scaling. To find out how this measurement
can then be used to learn about cluster physics, we compare these
findings with the slopes and amplitudes for various ICM physics models
according to \citet{ReidSpergel}, hereafter RS. RS examine the
mass-SZ  scaling relations for several analytical and phenomenological
models of the intra-cluster medium, which span the range of plausible
models motivated by recent observations (see RS and references
therein). The approach presented here can of course be applied also to
other analytical models and to predictions from numerical work.
We show which ICM models we can distinguish based on these two
parameters.

\begin{figure} 
\includegraphics[width=\columnwidth]{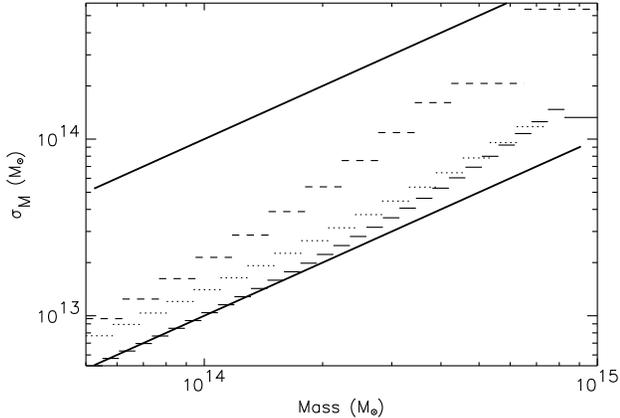}
\caption{Mass error ($\sigma_M$) vs. mass bin for three different redshift bins.  The dotted lines are for 
$0 < z < 0.2$, the thin solid lines are for $0.4 < z < 0.6$, and the dashed lines are for $1.1 < z < 1.3$.  The thick solid lines denote constant S/N: the lower, S/N=10; the upper, S/N=1. The error $\sigma_M$ has been doubled from the calculation based on the noise from galaxy shapes (sec. 2) in order to account for  the effect of large-scale structure along the line of sight.  The sky coverage is 200 sq. degrees, and the number density of galaxies is 20 per square arcminute, distributed as a Poisson distribution in redshift with mean $z=0.99$ ($z_0=0.33$). S/N will scale as the square-root of the sky coverage and as square-root of the number density of background galaxies.}
\label{fig:1} 
\end{figure}

We examine the mass range from $5 \times 10^{13} \, \Msun$ to $10^{15}
\, \Msun$ and consider 3 representative redshift intervals, $0 < z <
0.2$, $0.4 < z < 0.6$, and $1.1 < z < 1.3$.  We will hereafter refer
to these as redshift bins L (low), M (medium), and H (high),
respectively.  We choose the mass and redshift bins so that the error,
$\pm \sigma_M$, is greater than the size of the bin, $\pm \frac{\Delta
M}{2}$, and we have more than one cluster per bin.  Note that the three redshift bins 
are disjoined, therefore for a fixed survey area stronger constraints can be obtained by considering additional redshift bins and/or combining together all redshifts. We choose the sky
coverage to be 200 square degrees, and $n_g=20$ source galaxies per
square arcminute from a Poisson redshift distribution with $z_0= 0.33$
(mean redshift 0.99).  As the reported errors scale with
$1/\sqrt{n_g}$ and $1/\sqrt{f_{sky}}$, quadrupling the sky coverage or
the galaxy density would halve the errors.  Figure \ref{fig:1} shows
the mass bins chosen and the error per bin\footnote{Note that the
error per bin depends on the bin size, bin sizes at a given redshift
are approximately constant in log $M$ except for the highest masses,
but binsizes differ between different redshift bins.} for the three
redshift bins and two lines of constant S/N.  The mass bin sizes and
the number of clusters per bin are reported in tables
\ref{table:numgals1}, \ref{table:numgals2}, and \ref{table:numgals3}.
In the figures we have included large-scale structure effects by
doubling the errors (pessimistic case).

Of course, mass cannot be directly observed.  Forthcoming experiments
(e.g., ACT, SPT) will observe the integrated SZ effect, $L_{sz}$.  In
practice, therefore, clusters will be binned by their $L_{sz}$.  From
numerical simulations (e.g., \citet{Motl,Nagai}) the $M$-$L_{sz}$
relation is expected to be monotonic, $L_{sz}=A (M/M_o)^{\alpha_M}$,
with small (\% level) scatter, enabling one to assign an average mass
to a small $L_{sz}$ bin without introducing additional
scatter\footnote{The reader may note here that the scatter in the
M-$L_{sz}$ relation can introduce something similar to Malmquist bias
as lower mass clusters are more numerous than higher mass ones. Once
the width of the scatter in the M-$L_{sz}$ relation is known this effect can be
taken into account. For our purposes, given the small scatter found in simulations, we neglect this (small) effect.}.  One can of course test this assumption a
posteriori by comparing the predicted mass error with the intrinsic
{\it r.m.s.} in a given bin for bins with high enough S/N.

Here we assume a fiducial relationship between mass and $L_{sz}$, with
$\alpha_M = 1.66$ and $A = 6.18 \times 10^{-6}$ Mpc$^2$ at $M_o =
1.39 \times 10^{14} \Msun$, corresponding to a self similar model, without
non-gravitational physics (i.e. heating and cooling) where e.g. the 
gas temperature is given solely by the dark matter virial temperature.  

\begin{figure} 
\includegraphics[width=\columnwidth]{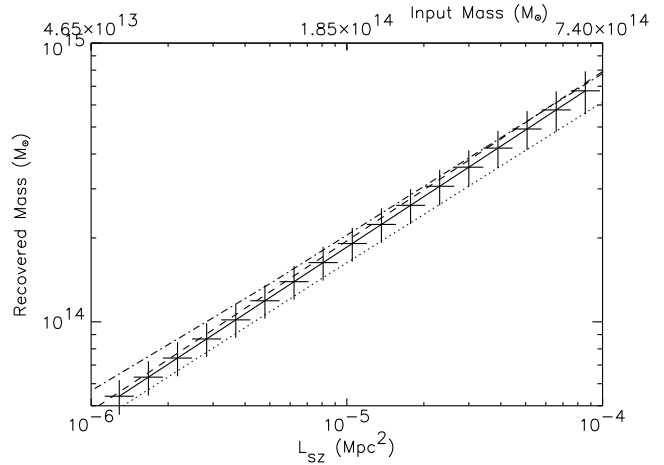}
\caption{{\it Left}: Recovered mass vs. input  mass and $L_{sz}$ for redshift bin L: $0 < z < 0.2$.  The horizontal lines denote the mass
bin and the vertical lines denote the error on the average mass in
that bin. {\it Right}: Using the self-similar model from \citet{ReidSpergel} as a fiducial model (solid line), we have transformed each mass bin into a
bin in $L_{sz}$, and again plotted the bin width and error as crosses.
The dotted-dashed line shows model 1, the dashes line shows model 2, and the dotted line shows model 3. Note that the slope in this plot is $1/\alpha_M$.}
\label{fig:2} 
\end{figure}

\begin{figure} 
\includegraphics[width=\columnwidth]{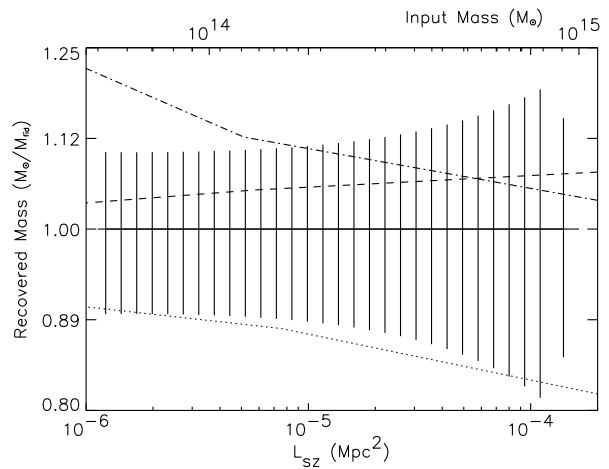}
\caption{{\it Left}: Recovered mass vs. input mass and $L_{sz}$ for redshift bin M: $0.4 < z < 0.6$.  The horizontal lines denote the mass
bin and the vertical lines denote the error on the average mass in
that bin. {\it Right}: Using the RS self-similar model as a fiducial model, we have transformed each mass bin into a
bin in $L_{sz}$, and again plotted the bin width and error as crosses.
The dotted-dashed line shows model 1, the dashes line shows model 2, and the dotted line shows model 3. Note that the slope in this plot is  $1/\alpha_M-1/\alpha_{M}^{\rm FIDUCIAL}$.}
\label{fig:3} 
\end{figure}

\begin{figure} 
\includegraphics[width=\columnwidth]{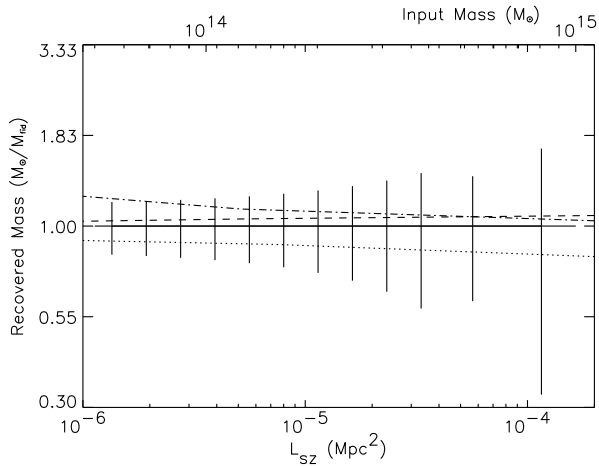}
\caption{{\it Left}: Recovered mass vs. input mass and $L_{sz}$ for redshift bin H: $1.1 < z < 1.3$.  The horizontal lines denote the mass
bin and the vertical lines denote the error on the average mass in
that bin rescaled by the fiducial. {\it Right}: Using the RS self-similar model as a fiducial model, we have transformed each mass bin into a
bin in $L_{sz}$, and again plotted the bin width and error as crosses.
The dotted-dashed line shows model 1, the dashes line shows model 2, and the dotted line shows model 3. Note that the slope in this plot is  $1/\alpha_M-1/\alpha_{M}^{\rm FIDUCIAL}$.
}
\label{fig:4} 
\end{figure}

Using this fiducial model, we thus transform the mass bins into bins in
$L_{sz}$.  The results are shown in figures \ref{fig:2}, \ref{fig:3},
and \ref{fig:4} for mass bins L, M and H respectively.  We then compare these scaling relations and their
errors with the theoretical models worked out by \citet{ReidSpergel},
hereafter RS.

Our fiducial model corresponds to their self-similar phenomenological
model.  RS use a different equation for the concentration parameter
$\C(M,z)$, but since we average over a range of $\C$, it amounts to a
small effect which we neglect here.  Their models focus on the entropy
parameter $K = P/\rho^{5/3}$, where $P$ is pressure and $\rho$ is the
density of the gas, so the gas entropy is proportional to $\ln K$.
The gas entropy profile is parameterized by a normalization
($K_\mathrm{max}$) and a radial profile with a power law index ($s_1$), and optionally
a core with smaller power law index. Cooling and/or preheating affect
$K_\mathrm{max}$ and the profile.
Preheating the gas causes it to start with a greater ``preheating''
entropy $K_\mathrm{ph}$, which serves to help offset the radiative
cooling.  Models of radiative cooling can reproduce important features of trends in X-ray clusters.  However, including only radiative cooling cooling produces too much cooled gas, so some feedback is necessary. Other models deal with varying the
accretion pressure.  In the RS analytical models, the pressure is
computed from the ram pressure of the infalling gas.  The accretion is
assumed to be smooth and spherical, while in reality is lumpy and stochastic. Deviations from these
assumptions would change the accretion pressure.  RS also examine
models whose entropy profiles are derived by assuming a simple
polytropic equation of state, $P \propto \rho^\gamma$, with pre-shock
gas density $\rho_1$ and post-shock gas density $\rho_2$ (the gas is
shocked as it accretes onto the cluster, when it crosses the sound
barrier).

Our fiducial model corresponds to RS self-similar model with an entropy profile $K \propto r^{1.1}$, a baryon fraction of 0.13, an entropy normalization $K_\mathrm{max} = K_{100}$.
The two models with extreme values of $\alpha_M$ are the RS analytical
model with the lowest amount of preheating, $K_\mathrm{preheat} =
10^{30}$ erg cm$^2$ g$^{-5/3}$ (and the largest fraction of cooled
gas, $f_\mathrm{cool}=0.42$), which we will call model 1, and the
phenomenological model with an accretion pressure decreased by a
factor of 3.5 from self-similar spherical collapse, model 2.  The
models with extreme values of amplitude $A$ are model 1 and the
phenomenological model with an accretion pressure increased by a
factor of 3.5 from self-similar spherical collapse, model 3. Table
\ref{table:slopeint} shows the best-fit slope, $\alpha_M$,  and amplitude, $A$, for the $M$-$L_{sz}$ relation for these
models.
We have included these three
models in the figures, where the dotted-dashed line denotes the low
preheating value (model 1), the dashed line denotes the lower
accretion pressure (model 2), and the dotted line denotes the higher accretion pressure (model 3). 
Performing a joint analysis using both the slope and and the amplitude
reveals that we can distinguish models 1 and 3 at the $ \geq 3\sigma$ level in the low
redshift bin, and at better than 5 $\sigma$ in redshift bin M.  For the
survey parameters and redshift bins chosen, we cannot distinguish model 2
very well in any redshift bin (see table \ref{table:joint}).  However, if we combine measurements from all 
redshifts or  if we increased the size of the survey to 1000 sq. deg. and, in this case, consider only redshift bin M, we could distinguish it at the $3\sigma$ level.

\begin{table}
\begin{ruledtabular}
\begin{tabular}{ccc}
Cluster model & $\alpha_M$ & $A$ (kpc$^2$) \\
Fiducial (self-similar) & 1.66 & 6.18 \\
Model 1 & 1.77 &  4.76\\
Model 2 &  1.64 & 5.74 \\
Model 3 & 1.71 &  7.78\\
\end{tabular}
\end{ruledtabular}
\caption{The best-fit slopes $\alpha_M$ and amplitudes $A$ for RS
models plotted in our figures, where $L_{sz}=A (M/M_o)^{\alpha_M}$ and
$M_o=10^{14}\Msun/h$.
Model 1 is the RS model with preheated entropy parameter
$K=10^{30}$ erg cm$^2$ g$^{-5/3}$, model 2 is the RS model with decreased
accretion rate $\omega_{\mathrm accr}=1/3.5$, and model 3 is the RS model with increased accretion rate $\omega_{\mathrm accr}=3.5$.
}
\label{table:slopeint}
\end{table}

\begin{table*}
\begin{ruledtabular}
\begin{tabular}{cllllll}
Cluster model &   & & & &  & \\
 \multicolumn{7}{c}{\hspace{0.7in} \# $\sigma$ from fiducial, pessimistic \hspace{0.5in}   \# $\sigma$ from fiducial optimistic}\\
 \hline
 &  $z$ bin L & $z$ bin M & $z$ bin H & $z$ bin L & $z$ bin M & $z$ bin H \\
 $K_\mathrm{ph} = 10^{30}$ (Model 1)&   3&$>$5& 1& 3.5 &$>$5  &1.5\\ 
 $\omega_\mathrm{accr}=0.28$ (Model 2)&   $0.5$&1.3 &$<$1  & $~$1&2&$<$1 \\ 
 $\omega_\mathrm{accr}=3.5$ (Model 3)&   3.5 & $>$5& 1& 4& $>$5 &1.3 
\end{tabular}
\end{ruledtabular}
\caption{The number of $\sigma$ from the fiducial to which the RS models
plotted in our figures can be distinguished, using a joint analysis
with both slope and amplitude, in the three redshift bins and in both
the optimistic and pessimistic cases.}
\label{table:joint}
\end{table*}

We expect that some systematic effects may affect much more one parameter than the other. If this is the case, one may want to consider the constraints on only  one parameter  (marginalized over the other one).  Therefore, next we look
at what  can be learned using  just slope or just amplitude
information.

For the slope,  we find that in the
pessimistic case the error on the slope is
$\sigma_{\alpha_{M}}=0.095$ for bin L,
$\sigma_{\alpha_{M}}=0.052$ for bin M
and $\sigma_{\alpha_{M}}=0.24$ for bin H.

We can thus distinguish model 1 from the fiducial to about $2\sigma$ in redshift bin M, but cannot distinguish much else in this case from slope information alone (see table \ref{table:Lszslope}). 

Since the error in the
slope is so small for redshift bin 2, due to the good signal to noise
(and therefore large number of mass bins allowed), we can
also distinguish among closer RS models in this case.  For example, in the pessimistic case, we can
distinguish between the two phenomenological models with varying
accretion (by factors of 3.5 and 1/3.5) to more at about $2\sigma$, and we can distinguish between the two extreme slopes of the RS
preheating models to more than $1\sigma$. \citet{Nagai} found $\Delta \alpha_M \sim 0.1$ between a purely
adiabatic cluster model and one that included cooling and star
formation.  Clearly in redshift bin M  for a  sky coverage $> 200$ square degrees, we can begin to distinguish differences of this order and thus discriminate between these models.  

Both analytical and numerical work (e.g, \citet{Nagai,ReidSpergel, Motl, McCarthy03})
show that different cluster physics modifies the amplitude of the
$M-L_{sz}$ scaling much more than the slope.  Nagai finds a difference
of about 30\% in amplitude between his purely adiabatic simulation and
one with cooling and star formation.  RS models differ by up to 90\%
in amplitude at $M=10^{14} \Msun / h$.

  Thus we are very interested in how well we can constrain the
amplitude of this relation.  To do this, the weak lensing
determination of clusters masses need to be accurately calibrated to
yield the clusters virial masses. Here first we assume that this can
be achieved with a negligible residual bias, although this is indeed
an assumption whose validity remains to be tested. Then we show how
relaxing this assumption (i.e. allowing a residual systematic error)
degrades the results.

Assuming gaussian errors and marginalizing over the slope, we find the
$1\sigma$ constraints on the amplitude $A$ of the $M-L_{sz}$ relation to be 
$5.67 - 6.55 \times 10^{-6}$ Mpc$^2$ 
for redshift bin L, 
$5.85 - 6.34 \times 10^{-6}$ Mpc$^2$ 
for redshift bin M,
and
$5.3 - 7.0 \times 10^{-6}$ Mpc$^2$
for redshift bin H.

Using amplitude alone, we can distinguish models 1 and 3 from the fiducial at better than $3\sigma$, especially in the M redshift bin (see table \ref{table:Lszamp}).
In the pessimistic case, we can
distinguish all of the preheating models (from $K=5 \times 10^{33}$ to
$K=10^{30}$) from the fiducial to more than $2\sigma$ in redshift bins L
and M.

Up to now we have assumed that the calibration of the mass
determination has negligible residual systematic error. A systematic
uncertainty in the mass determination of $x$\% would propagate into
$\sim 1.66\times x$\% systematic in $L_{sz}$ to be added to the
uncertainties in the amplitudes reported above. Thus, for example, $x=5$
would halve (reduce by a factor 2.5) the number of $\sigma$s reported for
bin M.

As pointed out by, e.g., \citet{ReidSpergel, Motl}, the scaling of the
central Compton parameter, $y_0$, with mass is far more sensitive to
the cluster model than the integrated $L_{sz}$.
The effect of cluster physics on the $L_{sz}-y_0$ relation has been studied by e.g. \citet{Mccarthy+03}. Here
instead we investigate what additional information can be learned from
measuring $y_0$ in addition to $L{sz}$ and weak-lensing mass determination. Even with a single model, however,
there is a large scatter in the $M-y_0$ relation \citep{Motl}.  This
partly counter-balances the advantage that the slopes for different models
are farther apart. 

As argued above, clusters could be binned in $L_{sz}$, and the average
mass per bin computed from the stacked clusters using a fiducial
model. Then one could measure the $y_0$ for each cluster in the
bin and take the average. This will have an additional error
associated because of the intrinsic scatter in the $M-y_0$ relation.

\begin{figure} 
\includegraphics[width=\columnwidth]{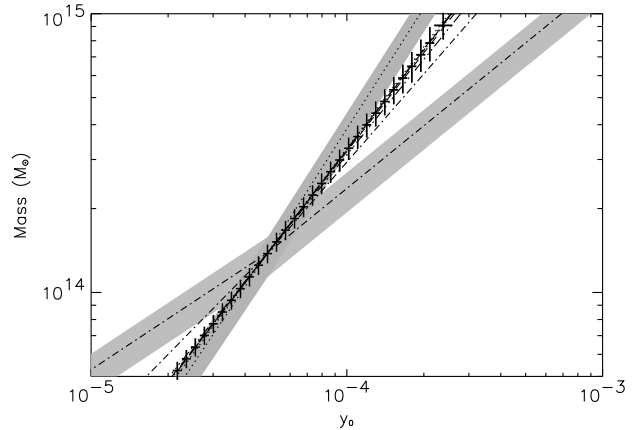}
\caption{Cluster physics has a bigger effect on the the scaling of mass with $y_0$.
This plot shows assorted lines from RS for the mass-$y_0$ scaling,
with amplitudes normalized to agree at $M=10^{14} \Msun/h$.  The
crosses show the mass bins and errors for the fiducial model, the RS
self-similar phenomenological model, for redshift bin M.  The gray
regions show the intrinsic scatter in $y_0$ according to \citet{Motl}
for the two most extreme slopes, models 4 (dotted-dashed) and 5 (dotted).}
\label{fig:y0} 
\end{figure}

\begin{figure} 
\includegraphics[width=\columnwidth]{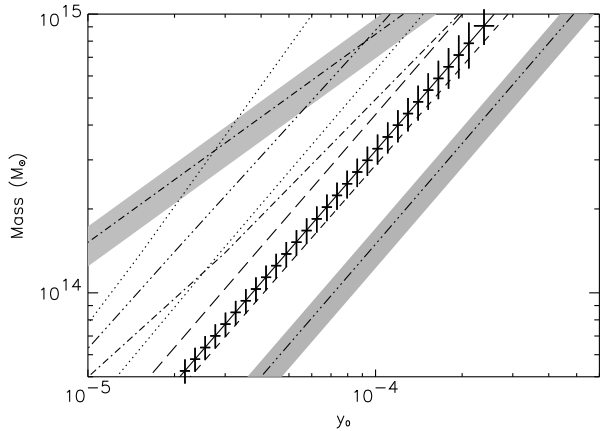}
\caption{ Assorted models from RS for the mass-$y_0$ scaling, with
amplitudes not normalized.  The crosses show the bins and errors for
the fiducial model, the RS self-similar phenomenological model, for
redshift bin M.  The gray regions show the intrinsic scatter in $y_0$
according to \citet{Motl} for the two most extreme amplitudes.  The dotted-dashed lines show the extreme preheating models, the dotted lines show the extreme polytropic models, the dashed lines show the extreme accretion pressure models, and the triple-dotted dashed lines show the different values for $s_1$, the exponent of the radius in the radial profile of $K$.}
\label{fig:y02} 
\end{figure}

Taking the RS self-similar model as our fiducial, 
with a slope $0.84$ and an amplitude $4.92$, 
we plot the expected
bin sizes and mass errors for redshift bin M as crosses on the
fiducial line (see figure \ref{fig:y0}).  We show the expected scatter from \citet{Motl}, who computed the scatter from simulations of about 100 clusters, as a shaded gray region around the extreme models.
Note that for a
survey of 200 square degrees, one expects to detect about 2000 clusters,
so the effect of this scatter on the determination of slope and
amplitude of the scaling relation should be reduced.   
The two RS
models with extreme slopes for this scaling are the analytical model
with the largest preheating ($K_\mathrm{preheat}=5 \times 10^{33}$ erg
cm$^2$ g$^{-5/3}$), model 4, and the polytropic model with $\gamma
=1.2, \rho_2=1.06 \rho_1$, model 5.  We can distinguish model 4's
slope from the fiducial model at more than the $5\sigma$ level for all
redshift bins, and can distinguish model 5 quite well in the lower two redshift bins (see table \ref{table:yamp}). 

Figure \ref{fig:y02} shows the lines for various RS models without
normalized amplitudes.  The models with extreme amplitudes are model 4 (the preheating model)
and model 6, the model with \footnote{Recall that $s_1$ parameterizes the entropy profile,  $K \propto r^{s_1}$ and
that $s_1=1.1$ 
  for the fiducial model.} $s_1=1.5$.  We can distinguish both models'
amplitudes from the fiducial model at $\gg$ 5 sigma in all redshift
bins and we can start constraining the parameters describing cluster physics, e.g., the amount of entropy injection.

Table \ref{table:yamp} shows a few RS models, their amplitudes for the
$y_0-M$ scaling, and their $\sigma$ from the fiducial.  Notice we can
easily distinguish models with amplitudes that vary by about 20\%.
Physics that causes such a difference in amplitude includes changing
the entropy profile normalization by a factor of 1.5, doubling the
radius at which the entropy maxes out, changing the accretion rate by
a factor of 3.5, or increasing the exponent $s_1$ of the radial entropy
profile from 1.1 to 1.5.

As the slope for the $y_0$ relation is 0.84
(range 0.71 - 1.5), an additional  $x$\% systematic error on the mass determination would propagate into an additional $\sim 0.8\times x$\%   
error on the amplitude of the scaling relation.  Even allowing for a
10\% systematic mass error we find that, for redshift bin M, we can
distinguish all of the RS preheating models, polytropic models,
$s_1=0.7$, and $s_1=1.5$, to more than $5\sigma$.  We can distinguish
the models with varying accretion pressure (models 2 and 3) to more
than $2\sigma$.  The range of amplitudes $A_{y}$ are $7.0 \times
10^{-6}$ to $9.4 \times 10^{-5}$ (fiducial $4.9 \times 10^{-5}$).

Since the models with maximum and minimum slope and amplitude are different
for $y_0$ than for $L_{sz}$, combining the two would allow for further
distinctions among models of cluster physics.

In the cases where the two parameters fit yields a $S/N >5$ in
principle there could be enough S/N to add parameters to the fit, or
in other words to investigate if e.g., deviations from a power law in
the scaling relation can yield additional information about cluster
physics.

Here we have concentrated on a survey with the properties and sky
 coverage of ACT and its optical follow up.  Naturally a wider SZ
 survey with weak lensing follow up and/or a weak lensing follow up
 from space will improve these constraints.  A space telescope can
 yield $n_g=100$ galaxies per square arcminutes. While for low and
 intermediate redshift this will improve constraints reported here
 only by $~\sqrt{5}$, as the additional galaxies are likely to have a
 distribution with a higher-redshift tail, the constraints at the
 higher redshift will show a better improvement. This may open up the
 possibility to study redshift evolution of the ICM. On the other hand
 a ground-based survey covering 1000 square degrees will also yield
 constraint better by a factor $\sqrt{5}$, providing better discriminative  
power between models. Finally a combination of ground-based SZ
 observations over 1000 square degrees with space-based weak lensing
 follow up would yield improved constraints by a factor 5. This combination is not too dissimilar from the planned SNAP observations\footnote{snap.lbl.gov} if ACT could scan in SZ the same  region of the sky.

\begin{table*}
\begin{ruledtabular}
\begin{tabular}{ccllllll}
Cluster model &  $\alpha_{M}$ & & & & &  & \\
 \multicolumn{8}{c}{\hspace{1.0in} \# $\sigma_{\alpha_M}$ from fiducial, pessimistic \hspace{0.5in}   \# $\sigma_{\alpha_M}$ from fiducial optimistic}\\
 \hline
 & & $z$ bin L & $z$ bin M & $z$ bin H & $z$ bin L & $z$ bin M & $z$ bin H \\
$K_\mathrm{ph} = 10^{30}$ (Model 1)& 1.77 &  1.1&2& 0.4&1.4&2.6  &0.6\\ 
$\omega_\mathrm{accr}=0.28$ (Model 2)&  1.64 &  0.5&1 &0.2  & 0.7&1.3&0.3 \\ 
\end{tabular}
\end{ruledtabular}
\caption{
Two extreme slope models we consider, described by \citet{ReidSpergel}, their
best-fit slopes $\alpha_M$ (where $L_{sz}=A (M/M_o)^{\alpha_{M}}$), and the number of $\sigma$ between their slopes and the fiducial slope in the three redshift bins and in both the optimistic and pessimistic cases.
}
\label{table:Lszslope}
\end{table*}

\begin{table*}
\begin{ruledtabular}
\begin{tabular}{ccllllll}
Cluster model &  $A$ (kpc$^2$) & & & & &  & \\
 \multicolumn{8}{c}{\hspace{1.3in} \# $\sigma_{A}$ from fiducial, pessimistic \hspace{0.5in}   \# $\sigma_{A}$ from fiducial optimistic}\\
 \hline
 & & $z$ bin L & $z$ bin M & $z$ bin H & $z$ bin L & $z$ bin M & $z$ bin H \\
$K_\mathrm{ph} = 10^{30}$ (Model 1)&  4.76   & 3 & 6  &  1.8 & 4.5& 8 & 2.3
\\ 
$\omega_\mathrm{accr}=3.5$ (Model 3)&   7.78 & 3& 5& 1.6& 3.9&7 &2
\\  
\end{tabular}
\end{ruledtabular}
\caption{
Two extreme amplitude models we consider, described by \citet{ReidSpergel}, their
best-fit amplitudes $A$  (where $L_{sz}=A (M/M_o)^{\alpha_{M}}$), and the number of $\sigma$s between their amplitudes and the fiducial amplitude in the three redshift bins and in both the optimistic and pessimistic cases.
}
\label{table:Lszamp}
\end{table*}

\begin{table*}
\begin{ruledtabular}
\begin{tabular}{ccllllll}
Cluster model & $\alpha_{My}$  & & & & &  & \\
 \multicolumn{8}{c}{\hspace{1.5in} \# $\sigma_{\alpha_{My}}$ from fiducial, pessimistic \hspace{0.3in}   \# $\sigma_{\alpha_{My}}$ from fiducial optimistic}\\
 \hline
  & & $z$ bin L & $z$ bin M & $z$ bin H & $z$ bin L & $z$ bin M & $z$ bin H \\
\tablecaption{Two extreme slope models (for the central Compton parameter) we consider, described by \citet{ReidSpergel}, their
best-fit slopes $\alpha_{My}$ (where $y_0=A_y (M/M_o)^{\alpha_{My}}$), and the number of $\sigma$ between their slopes and the fiducial slope in the three redshift bins and in both the optimistic and pessimistic cases.}
$K_\mathrm{ph}=5 \times 10^{33}$ (Model 4) & 1.50   & $>$5 &$>$5  &$>$5   &$>$5  &$>$5  & $>$5    \\ 
$\gamma = 1.2$, $\rho_2=1.06 \rho_1$ (Model 5)  & 0.71  & 2&3&1&3&5&1.2 \\
\end{tabular}
\end{ruledtabular}
\caption{Two extreme slope models (for the central Compton parameter) we consider, described by \citet{ReidSpergel}, their
best-fit slopes $\alpha_{My}$ (where $y_0=A_y (M/M_o)^{\alpha_{My}}$), and the number of $\sigma$ between their slopes and the fiducial slope in the three redshift bins and in both the optimistic and pessimistic cases.}
\label{table:yslope}
\end{table*}

\begin{table*}
\begin{ruledtabular}
\begin{tabular}{ccllllll}
Cluster model & $A_y$ ($10^{-5})$  & & & & &  & \\
 \multicolumn{8}{c}{\hspace{1.7in} \# $\sigma_{A_y}$ from fiducial, pessimistic \hspace{0.3in}   \# $\sigma_{A_y}$ from fiducial optimistic}\\
 \hline
 & & $z$ bin L & $z$ bin M & $z$ bin H & $z$ bin L & $z$ bin M & $z$ bin H \\
$K_\mathrm{max}=K_{200}$  (Model 7)  &5.66&2.9&4.4&1.9&3.2&4.7&2.2\\
$K_\mathrm{max}=1.5 K_{100}$ (Model 8)&3.84&5  &$>$6 &3.4&5.6&$>$6 &4\\
$\omega_\mathrm{accr}=0.28$ (Model 2)&5.41&1.9&3  &1.3&2.1&3.1&1.5   \\
$\omega_\mathrm{accr}=3.5$ (Model 3) &4.33&2.7&4  &1.8&2.9&4.3&2    \\
$K_\mathrm{ph}=5 \times 10^{33}$ (Model 4) &0.70   & $>$6 &$>$6  &$>$6   &$>$6  &$>$6  & $>$6\\ 
$s_1=s_2=1.5$ (Model 6) & 9.39& $>$6 &$>$6  &$>$6   &$>$6  &$>$6  &$>$6\\
\end{tabular}
\end{ruledtabular}
\caption{Sample models for the central Compton parameter that fall within 2-6 $\sigma$ from the fiducial in amplitude, described by \citet{ReidSpergel}, their
best-fit amplitudes $A_y$  (where $y_0=A_y (M/M_o)^{\alpha_{My}}$), and the number of $\sigma$ between their amplitudes and the fiducial amplitude in the redshift bins and in both the optimistic and 
pessimistic cases.}
\label{table:yamp}
\end{table*}

\section{Discussion and Conclusions}

We have shown how, stacking the weak lensing signals from multiple
clusters with roughly the same SZ luminosity, the SZ luminosity-mass
($L_{sz}-M$) and SZ central decrement-mass ($y_o-M$) relations can be
measured from forthcoming SZ surveys with extensive optical follow
up. For example, from a survey of 200 square degrees such as ACT, we
find we should be able to determine the scaling slope to $\pm
0.05-0.2$ and amplitude to 5-20\% depending on redshift bin.
Observations indicate that the entropy of the gas in clusters cannot
be explained by gravitational collapse alone, but the cause of the
increased entropy remains to be determined.  Different
non-gravitational processes affect both the slope and amplitude of
these relations, and in particular the $y_o-M$ relation is most
sensitive to cluster physics.  Used in conjunction with results from
simulations, the expected errors on the measurements of the slope and
amplitude of scaling relations imply that we can start to discriminate
among different non-gravitational processes affecting the ICM.

We have shown that, for a survey of $\sim$ 200 sq. deg., the available
S/N enables one to measure these scaling relations as a function of
redshift: we can distinguish slopes and amplitudes of the $M-L_{sz}$
relation and $M-y_0$ relation for different redshift bins with width
$\Delta z = \pm 0.1$, thus opening up the possibility to constrain how
the mass-SZ scaling evolves with redshift. A mass
measurement from weak lensing and a SZ measurement in different
redshift bins,  can constrain the evolution of hot gas as a function
of redshift, which in turn would enable one to constrain feedback
evolution.  Inspection of Fig. \ref{fig:3} and \ref{fig:4} provides a
quantitative measurement of how physics in clusters, e.g. feedback,
can be constrained as a function of redshift. The errors
reported here will scale with the survey parameters as $1/\sqrt{f_{\rm
sky}}$ and $1/\sqrt{n_g}$.  Thus, using this method, wider surveys can
improve constraints on cluster physics just as well as deeper surveys.
For a space telescope with $n_g=100$ galaxies per
square arcminute and 1000 square degrees, our results would improve by
a factor of 5.

A calibration of the SZ-Mass relation with known uncertainties 
also has consequences for the determination of cosmological parameters.  The
number density of clusters within a given mass range or above a given
mass is extremely sensitive to cosmology. This sensitivity, in
principle, can lead to tight constraints on cosmological parameters
such as $\Omega_m$, dark energy properties and neutrino mass.
However, even relatively small systematic errors in clusters mass
estimates can lead to systematic errors on the recovered cosmological
parameters that are larger than the statistical ones \citep{HolderHaimanMohr, Battye03,Francis05}.
This potentially compromises measurements of parameters like e.g. the dark
energy equation of state. The technique outlined in this paper to
calibrate the mass-SZ relation may help reduce the amplitude of these
systematic errors. The weak lensing mass determination used here is
sensitive to the mass along the line of sight, not just the gas.
It is true that there can be biases introduced by the large-scale
structure local to the clusters, but we argue that this should not be
considered a fundamental limitation as this effect in principle can
be quantified by means of numerical N-body simulations. If this can be
achieved, our estimates above indicate that the residual uncertainty
in the calibration of the amplitude of the mass-SZ relation is $~ 5$\%
for a survey with $n_g=20$ galaxies per square arcminute and sky coverage of 200 square degrees.
This uncertainty would propagate into a systematic error in the
determination of cosmological parameters, that is now reduced to be 
not larger than the statistical uncertainty achievable from a survey of the same size (see \citet{Francis05}).  Alternatively, The M-L$_{\rm SZ}$ relation
calibrated in this way from a small area of the sky can be used to
determine masses of SZ clusters from very large SZ-only surveys, or the $M-y_0$ relation could be used to constrain cluster physics and this information can then be used to model the $M-L_{sz}$ relation.  
This approach is nicely complementary to other techniques proposed to
calibrate the mass-observable relations such as the self-calibration
using the cluster power spectrum of \citet{MajumdarMohr03}. While only
positions and redshifts of clusters are needed to measure the power
spectrum and thus to self-calibrate the SZ-Mass relation, it requires observations of large areas of the sky as square and
contiguous as possible. The calibration of the M-SZ relation with weak
lensing has more stringent follow up requirements (imaging with good
seeing in addition to clusters redshifts), but has less stringent
geometry requirements.

\section*{acknowledgments}
This research is supported in part by grant NSF AST-0408698 to the
Atacama Cosmology Telescope. RJ is partially supported by NSF grant
PIRE-0507768 and by NASA grant NNG05GG01G.  LV is supported by NASA
grants ADP03-0092 and ADP04-0093.  RJ and LV acknowledge the
hospitality of the Institute of Advanced Studies (Princeton) and
Institut d'Estudis Espacials de Catalunya (Barcelona) where part of
this work was done.


\end{document}